\documentclass[12pt]{amsart}
\usepackage{amsmath,amsfonts,amssymb,amscd,amsthm,epsf}
\usepackage{graphics}
\newtheorem{theorem}{Theorem}[section]
\newtheorem{proposition}[theorem]{Proposition}

\theoremstyle{definition}

\newcommand{\bbR}{{\mathbb R}}
\newcommand{\bbZ}{{\mathbb Z}}
\newcommand{\bbC}{{\mathbb C}}

\newcommand{\be}{\begin{equation}}
\newcommand{\ee}{\end{equation}}

\theoremstyle{remark}

\begin{document}
\title{Semiclassical analysis and sensitivity to initial conditions}
\footnote{Talk given at the Colloquium in Honour of Giuseppe Longo, Paris June 28-29, 2007}
\author{Thierry Paul}
\address{CNRS, D\'epartement de Math\'ematiques et Applications UMR 8553}
\email{paul@ dma.ens.fr}
\address{Ecole Normale Sup\'erieure, 45, rue d'Ulm - F 75730 Paris cedex 05}

\maketitle

\hfill{\it pour Giuseppe\hskip 1.2cm}

\date{}

\begin{abstract}
We present several recent results concerning
the transition between quantum and classical mechanics, in the situation
where the underlying dynamical system has an hyperbolic behaviour. The
special role of invariant manifolds will be emphasized, and the long time
evolution will show  how the quantum non-determinism and the classical
chaotic sensitivity to initial conditions can be compared, and in a certain
sense overlap. 
\end{abstract}

\vskip 1cm

\tableofcontents

\section{Introduction}
Quantum Mechanics has deeply changed our vision of the world. Non-relativistic Classical Mechanics involves very strongly the notion of  absolute space, in which the concept of ``material point"  
is present, obviously as an idealization. This space is
supposed to be independent of the objects it contains, and moreover independent of their dynamics. Quantum Mechanics presents a different situation, as we shall explain 
in details later on, where the classical notions of points, trajectories, intrinsic properties of objects, have to be dropped. Another paradigm is necessary, fundamentally
different and fully independent. Nevertheless, in order to ``see" (direct) quantum effects, we have to go in physics laboratories: our macroscopic vision of the world is
classical. A way of understanding this dichotomy is the following.  
There is a cursor which tells us how far, inside Quantum Mechanics, we are from Classical Mechanics: the Planck constant $\hbar$. As $\hbar\to 0$ the classical boundary of
Quantum Mechanics is reached and most of quantum effects disappear. Most of them, but not all of them, especially when one considers long time behaviour. In this paper we would
like to present recent results stressing this classical or non-classical limiting procedure.

Before we summarize the results developed in this paper, let us try to mention several common features that  Quantum Mechanics and Computer Science share. Computer Science
has also changed our vision of the world, or at least the way we have to figure out a model of it. Let us take the example of numerical methods in applied mathematics. 
The first thing that
a computer does when solving an equation is to discretize, in order to approximate the continuum by digital information. This discretization is also ``measured" by a certain
size, for example the number $N$ of digits kept in the decimal approximation. Larger is $N$, better is the approximation, and closer is the model to continuum. And the latter is
supposed to be reached when $N\to\infty$. Let us consider now that we mix this $N\to\infty$ together with a limit on the length of the algorithm the computer is running.
Obviously, for a good algorithm, at ``finite" time $L$ (length, number of steps) the continuum is reached back. But what happens (theoretically)  when $L$ has a
dependence on $N$, and $L(N)\to\infty$ as $N\to\infty$? Numerical computations in the theory of dynamical systems face exactly this problem: they digitize and compute in
finite time, and the conclusions they produce are interpreted inside continuous setting (for example strange sets, attractors etc) and infinite time (as the mathematical 
definition of a chaotic behaviour involves deeply the limit $t\to\infty$, as we shall comment later on).

Quantum Mechanics contains mathematics able to handle carefully this kind of problem, and we will see in this paper the surprise it creates, including ubiquity,
non-localisation, etc. We will focus on the link between an emblematic notion contained in the theory of ``chaos", namely the sensitivity to initial conditions phenomenon, and an
even more emblematic subject of Quantum Mechanics: the intrinsic undeterminism it contains. What we want to stress and will show in examples in this paper can be summarized this way:
\vskip 0.5cm

\noindent{\it as time $\to\infty$ quantum undeterminism and classical unpredictability merge.} 

\vskip 0.5cm

\section{A brief introduction to (infinite dimension) quantum mechanics}

The starting point in Quantum Mechanics is a (possibly infinite dimensional) Hilbert space $\mathcal H$ (complete normed space whose norm is given from a scalar product). The state of the system is given
by a vector $\psi$ in this space, and the evolution in time, outside measurement, is given by the Schr\"odinger equation:
\be\label{biq1}
i\hbar\partial_t\psi^t=H\psi^t
\ee
where $H$ is a self adjoint operator (densely) defined on $\mathcal  H$.

The self-adjointness property is required in order for the solution of (\ref{biq1}) to be obtained from the initial condition $\psi^0$ by the action of a unitary operator, that is
an operator preserving the norm defined on $\mathcal H$ 
\be\label{biq2}
\psi^t=U(t)\psi^0,\ U(t)=e^{-i\frac{tH}\hbar},\  ||\psi^t||=||\psi^0||
\ee

The fact that the norm is preserved by the quantum flow is the most elementary symmetry that  quantum mechanics possesses and is essential for the measurement process as we will see
now.

To a measurement is associated  (as in classical mechanics) a {\it quantum}\rm\  observable, which is a self-adjoint operator. The fact that an observable must be self-adjoint is required here
in order that the measurement corresponds to {\it real}\rm\  values. The spectral theorem ensures that we can associate to an observable $O$ a spectral decomposition given by, in the
case of discreetness of the spectrum,  a set on real numbers, the spectrum of $O$, $\sigma(O)=\{\lambda_j,\ i\in\bbZ\}$ and a family of orthogonal projectors 
 $\{\Pi_j,\ i\in\bbZ\}$ such that:
 \be\label{biq3}
 O=\sum_j\lambda_j\Pi_j
 \ee
 with also,
 \be\label{biq4}
 \sum_j\Pi_j=\mbox{Identity on}\ H.
 \ee
 The result of a measurement of the observable $O$ on the system in the state $\psi$ is given, randomly, by any number $\lambda_j$, with a probability
 $p(\lambda_j)=|\Pi_j\psi|^2$. Moreover the system is, right after the measurement  and in the case where the result was $\lambda_j$, in the state $\Pi_j\psi$.
 
 We see immediately that the condition on unitarity of the quantum flow is necessary in order to preserve in time the quantity:
 \be\label{biq5}
 \sum_jp(\lambda_j):=\sum_j|\Pi_j\psi|^2=|\psi|^2.
 \ee
 An immediate consequence of this is the possibility to define the so-called ``expectation value" of $O$ in the state $\psi$ as (we denote by $<.,.>$ the scalar product on
 $\mathcal H$): 
 \be\label{biq6}
 <O>:=<\psi,O\psi>=\sum_j\lambda_j|\Pi_j\psi|^2
 \ee
  which according to the interpretation of $|\Pi_j\psi|^2$ as a probability law, is the ``expectation" of O.
  
Physics likes very much differential equations,  a ``natural" choice of operators consists in differential operators. Therefore a ``natural" choice for $\mathcal H$ should be a
space of functions. Although there are plenty of Hilbert spaces of functions on $\bbR^n$, or more generally on a manifold, the space of square integrable functions is very
common (other choices with very great importance in quantum mechanics are spaces of analytic functions):
\[
L^2(\bbR^n,dx):=\{f: \bbR^n\to\bbC,\int_{\bbR^n}|f(x)|^2dx<+\infty\} 
\]
What is a differential operator? Roughly speaking it is an operator of the form $H=\sum a_l(x)\frac{d^l}{dx^l}$, that is an operator generated by the two operators $\times x$ and 
$\frac d {dx}$. We will see that there is a natural way of associating to a differential operator a function $f(p,q):=\sum_la^l(q)p^l$.

What is a point on a manifold? It can be seen as being a linear form on a space of functions defined on the manifold, the linear form of ``evaluation" of a function $f$ at the
point $x$. But evaluating a function is nothing but computing an expectation, with a singular probability law:
\be\label{biq7}
f(x)=\int f(y)\delta(x-y)dy
\ee
It is therefore natural to ask if there exists a family of vectors $\psi_{qp}$ in $\mathcal H$ such that $f(p,q)=<\psi_{qp},H\psi_{qp}>$. In general the answer is no, but it is almost true if we
introduce the Planck constant everywhere. The result is:
\begin{proposition}
There exists a family of vectors $\psi_{qp}$ such that, for any operator $H=\sum a_l(x)(-i\hbar\frac{d}{dx})^l$, we have, as $\hbar\to 0$,
\be\label{biq8}
<\psi_{qp},H\psi_{qp}>\to \sum a_l(q)p^l
\ee
\end{proposition}
 The vectors $\psi_{qp}$ are called ``coherent states".
 
 Let us finish this section by a remark. There is a way of synthesising quantum evolution in which the evolution is deterministic (Schr\"odinger equation), but what evolves is ...a probabilistic object: the amplitude of
probability. In this vision the Schr\"odinger equation is driving the fundamental quantity which will appear when a measurement is done. This last measurement part is fully
random, and the Schr\"odinger equation is here in order to let the probability evolve. It seems that, at a conceptual level, there is possibly here an analogy with the biological situation. Indeed biological
``evolution" is nowadays often seen as the influences of certain ``causes" \cite{BL} (for example created by the experimentalist) whose effects on the system is ``only" to modify, most of
the time by increasing, the variability of the system. Variability that will induce a fully random change of the system. The analogy is even more stringent if we notice the
phenomenon of ``spreading of wave packet" (actually at the heart of the main discussion on this paper): when evolving through the Schr\"odinger equation
 the wave function spreads around, and the
probability distribution increases its range, leading to  more variability for the results of measurement.

\section{The semiclassical limit}\label{scl}
\subsection{Coherent states and the concept of ``quantum point"}
Quantum mechanics deals with Hilbert spaces, and the state of the system is represented by a vector in such a Hilbert space. In the case where this Hilbert space is 
\begin{equation}
\mathcal H=L^2(\bbR^n,dx)\{f: \bbR^n\to\bbC,\int_{\bbR^n}|f(x)|^2dx<+\infty\}
\end{equation}
one can define the set of coherent states as the set of vectors minimizing the Heisenberg inequalities:
\begin{equation}\label{heis}
\Delta P\times\Delta Q\geq
\frac\hbar 2 
\end{equation}
where, for any selfadjoint operator $A$ and $\psi$ in the domain of $A$:
\begin{equation}
\Delta A:=\sqrt{<\psi,A^2\psi>-<\psi,A\psi>^2}
\end{equation}
(see \cite{TP}).
In the case of the canonical operators, $Q_i=\times x_i$ and $P_j=-i\hbar\partial_{x_j}$, an easy computation shows that the Heisenberg inequalities (\ref{heis}) are minimized by
the family of vectors:
\be\label{ec}
\psi_{qp}(x)=(\pi\hbar)^{-n/4}e^{-\frac{(x-q)^2}{2\hbar}}e^{i\frac{px}\hbar}
\ee
The arrow: $(q,p)\in\bbR^{2n}\longrightarrow
\psi_{qp}$ defines a one-to-one correspondence between $\bbR^{2n}$ and a subset (submanifold) of $\mathcal H$. We will see in this section how this is related to classical mechanics,
but let us for the moment consider that we have embedded in $\mathcal H$ a manifold $\Lambda_{sc}$ of vectors of $\mathcal H$ being the most classical as possible (in the sense
that they minimize the Heisenberg inequalities). It is obviously unthinkable to imagine that an initial condition ``pined-up" in this manifold should remain, after evolving, inside
the same manifold. What we are going to see is that this is nevertheless almost true. In fact what's going to be true is the fact that the evolved state will always be
decomposable as a sum of coherent states.

More precisely we will define, starting from $a\in\mathcal S(\bbR^n)$, the Schwartz class of functions on $\bbR^n$, and $(q,p)\in\bbR^{2n}$ the following vector
\be\label{eca}
\psi^a_{qp}=\hbar^{-n/4}a\left(\frac{x-q}{\sqrt\hbar}\right)e^{i\frac{px}\hbar}.
\ee
Therefore the canonical case of (\ref{ec}) corresponds to the Gaussian choice $a(\eta)=\pi^{-n/4}e^{-\frac{\eta^2}2}$.

\subsection{Time evolution}
The precise theorem of propagation of coherent sates reads as follows \cite{TP}:
\begin{theorem}\label{prop}
Let us consider the solution of the Schr\"odinger equation:
\be
i\hbar\partial_t\psi^t=H\psi^t
\ee
with an initial condition $\psi^{t=0}=\psi_{qp}$. Then, there exist $a_t,\ p(t),\ q(t)$ such that:
\be
\psi^t=e^{i\frac{l(t)}\hbar}\psi^{a_t}_{q(t)p(t)}+O(\hbar^{\frac 1 2 }))
\ee
Moreover $p(t),\ q(t)$ satisfy the following equation (with $^.$ means $\frac{\partial}{\partial t}$, that is, for example, $\dot q=\frac{dq}{dt}$)
\be\label{hami}
\left\{\begin{array}{ccl}
\dot q(t)&=&\frac{\partial h(q(t),p(t))}{\partial p(t)}\\
\dot p(t)&=&-\frac{\partial h(q(t),p(t))}{\partial q(t)}\\
\end{array}\right .,\ q(0)=q,\ p(0)=p.
\ee
The flow $\Phi^t:\ (q,p)\to(q(t),p(t))$ is called the Hamiltonian flow of Hamiltonian $h(q,p)$.
\end{theorem}
\subsection{The classical limit}
Taking the limit $\hbar\to 0$ in the preceding section leads to the principle of correspondence:
\begin{proposition}
Let $\mathcal O$ be an observable of the form given before, and let us consider the von Neumann equation, equivalent to the Schr\"odinger one,
\be\label{vn}
i\hbar\dot{\mathcal O}(t)=[\mathcal O(t),H]:=\mathcal O(t)H-H\mathcal O(t),\ \mathcal O(t=0)=\mathcal O,
\ee
then the symbol of $\mathcal O(t)$ is the one of $\mathcal O$ composed by the flow $\Phi^t$:
\be\label{cor}
\sigma(\mathcal O(t))=\sigma(\mathcal O)o\Phi^t.
\ee
\end{proposition}
But this is  a weak result compared to the propagation (of coherent states) Theorem \ref{prop}, in the sense that the propagation of states is much more precise, and reveals the
sensitivity to initial conditions, as we will see now.
\section{Classical mechanics and sensitivity to initial conditions}
As we mentioned in the introduction Classical Mechanics sits on the notion of absolute space, considered only at the ``receptacle" of the dynamics \cite{PC}. 
Nowadays  we always  forget
about the difficulty this notion had to emerge, as it seems so natural to us. In this section we want to discuss the way this vision is first of all a result of a limiting
procedure ($\hbar\to 0$) (remember the world is quantum), and secondly how, even by considering the classical situation as granted, it is an idealization, that we have to couple
with observations and (classical) measurement.
\subsection{Absolute space}
It is very familiar to us that Classical Mechanics is the theory of flows on spaces, with  certain symmetries. In the more general and ``fancy" formulation, the space is
a symplectic one, and the flow preserves this symplectic structure.
\subsection{Dynamics and Poincar\'e}
The classical flow, giving rise to trajectories, is usually given by the Hamiltonian equations, a system of $2n$ coupled ordinary equations of the form:
\begin{equation}\label{ham}
\left\{\begin{array}{ccl}
\dot q&=&\partial_ph(q,p)\\
\dot p&=&-\partial_qh(q,p)\end{array}\right.
\end{equation}
The symmetry preserved automatically by these equations is the symplectic form $dp\wedge dq$, and a corollary of this is the Liouville theorem which insure that the volume (of phase
space) is preserved.

For example, if the Hamiltonian $h$ is of the form $h(q,p)=\frac{p^2}2+V(q)$ the Hamiltonian equations take the form:
\begin{equation}\label{new}
\left\{\begin{array}{ccl}
\dot q&=&p\\
\dot p&=&-\partial_qV(q)\end{array}\right.\Leftrightarrow 
\left\{\begin{array}{ccl}
\dot q&=&p\\
\ddot q&=&-\partial_qV(q),\ \mbox{Newton equation}.\rm\end{array}\right.
\end{equation}

In general only the Hamiltonian function is preserved by the flow. The search for invariant integrals of motion, in involution, is the task of the theory of integrable systems.
When the number of such integrals of motion is maximal, that is equal to the number of degree of freedom, the system is called integrable, and its evolution fully understood: the
flow remains on (Lagrangian) tori, and is quasiperiodic (linear) on them. It was discovered by Poincar\'e that integrable systems are not stable by perturbations, giving rise to the
failure of a kind of Laplacian view of the world \cite{PO}.

But Poincar\'e took this lack of integrability much more seriously than a general negative result, and invented the theory of ``chaos", at the middle of which seats the concept
of sensitivity to initial conditions. Roughly speaking it says that, disregarding how close are two initial conditions, by waiting long enough, they are going to escape from each other
by a distance of any value. Mathematically it reads:
\begin{equation}\label{sci}
\exists I\in\bbR^+ , \forall\epsilon>0,\  \exists t=t(I,\epsilon)\ \mbox{such that}\ \exists x, |x-y|\leq\epsilon,\ |\Phi^t(x)-\Phi^t(y)|\geq I.
\end{equation}
Of course this definition involves  infinite time as, the flow being continuous, $t(I,\epsilon)\to\infty$ as $\epsilon\to 0$. We will comment later on this potential
infinite time, for which the concepts of unpredictability and determinism will overlap.

To get convinced that the sensitivity to initial conditions is something real let us consider the very simple case of an Hamiltonian flow of Hamiltonian function:
\be\label{ham23}
h(q,p)=qp
\ee
The Hamilton equations read:
\be
\left\{\begin{array}{ccc}
\dot q=q\\ \dot p=-p.\end{array}\right.
\ee
Therefore they can be easily integrated as:
\be
\left\{\begin{array}{ccc}
q(t)=e^tq_0\\  p(t)=e^{-t}p_0.\end{array}\right.
\ee
where $q_0,\ p_0$ is the initial condition. The important qualitative features of this example are the following:
\begin{itemize}
\item $q(t)-q'(t)=e^t(q_0-q'_0)$, exponential drift,
\item $(q(t),p(t))=(0,0),\ \forall t$ if $(q_0,p_0)=(0,0)$, fixed point at the origin,
\item $(q(t),p(t))\to (0,0)$ as $t\to +\infty$ if $q_0=0$
\item $(q(t),p(t))\to (0,0)$ as $t\to -\infty$ if $p_0=0$.
\end{itemize}
The two lines $\{(q=0,p)\}$ and  $\{(q,p=0)\}$ are called the \it{stable}\rm\ and \it{unstable}\rm\ manifolds of the fixed point $(0,0)$. We will see now that this situation is much
more general.
\subsection{Invariant objects}
The fact that a flow has the property of ``sensitivity  to initial conditions" has a deep impact on the possibility or impossibility of quantitative predictions, especially for long times.
Nevertheless there is a kind of miracle which provides to chaotic systems, in general, a very nice geometrical picture. This is the concept of stable and unstable manifolds,
as shown for example in Fig. 1.

\vskip 0.3cm

\centerline{\includegraphics{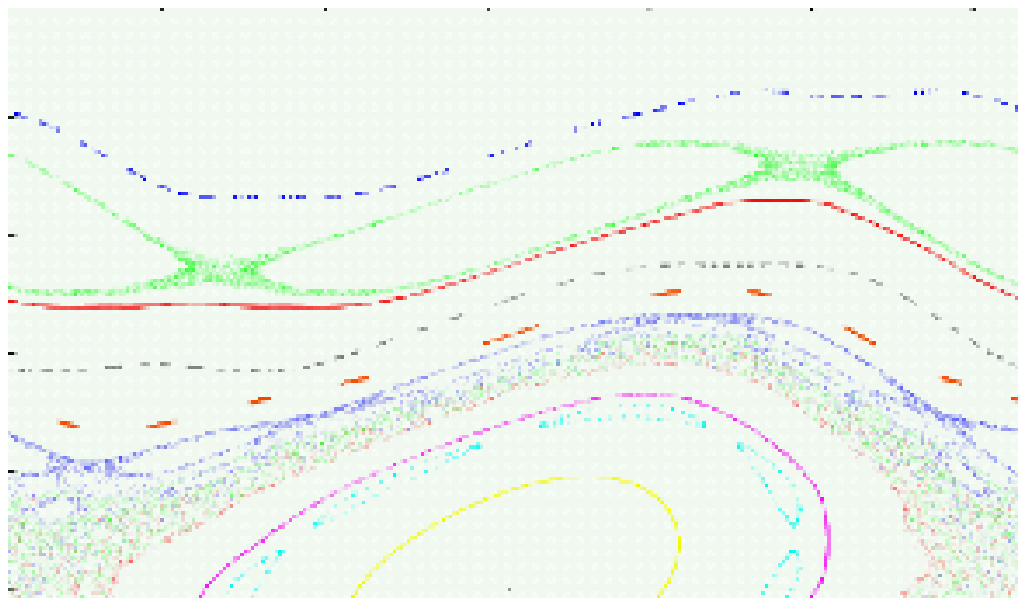}}

\centerline{Fig 1}

\vskip 0.3cm
Indeed let us consider again the condition (\ref{sci}) and, for sake of simplicity, let us suppose $y$ is a fixed point of the dynamics, $\Phi^t(y)=y\ \forall t$,
\be\label{sci2}
\exists I\in\bbR^+ , \forall\epsilon>0,\  \exists t=t(I,\epsilon)\ \mbox{such that}\ \exists x, |x-y|\leq\epsilon,\ |\Phi^t(x)-\Phi^t(y)|\geq I.
\ee
Considering the fact that, for autonomous flows, we have $\Phi^{-t}=\left(\Phi^t\right)^{-1}$, (\ref{sci2}) is equivalent to:
\be\label{sci3}
\exists X\ \mbox{such that}, \forall\epsilon>0,\exists t \ / \ |\Phi^{-t}(X)-y|\leq\epsilon
\ee
that is to say:
\be\label{sci4}
\exists X\ \mbox{such that:}\ \ \  \lim_{t\to-\infty}\Phi^t(X)=y=\Phi^t(y).
\ee
It turns out that the set of such $X$s has a very nice structure and is very important for studying chaotic flows.

In general let us  consider the set of
points associated to a given one $x$, that get closer and closer to its trajectory as $t\to\infty$,
\begin{equation}\label{inv}
\Lambda_x:=\{x'/
|\Phi^t(x')-\Phi^t(x)|\to 0\ \mbox{as}\  t\to\infty\}.
\end{equation}
As the first glance we could imagine that this set would not have any specific structure, and could be very complicated. It is complicated, but it has a very nice property: it is a
(embedded) Lagrangian manifold, Lagrangian in the same sense as for the tori of integrable systems.
\subsection{The ``seen" and the infinite time limit}
Trajectories of chaotic systems are very difficult to conceive independently each from the others. This is precisely the meaning, the deep meaning, of chaos. As two initial data
very close to each other are spread around (most of the time exponentially fast), and although two different trajectories will never meet, drawing trajectories is very
complicated, and doesn't give so much information. In fact what one usually draws in pictures in the theory of dynamical systems consists in packets of trajectories, precisely the
sets of trajectories which go around very well at the limit of long time. These trajectories are exactly the ones which form the stable and unstable manifolds we met in the last
paragraph (see again Fig. 1). What is really important, as far as infinite time evolution is concerned, are precisely the invariant objects we just constructed before.

But these objects, although they give a lot of information, and are, basically, the skeleton of the flow, are just considered in Classical Mechanics as tools, without any
ontological evidence. We will see later on how quantum mechanics gives them an ontological status.
\section{The case of dilations}
The case of the Hamiltonian $h=qp$ will give us a nice and simple example. Since the canonical quantization of ``q" is the operator of multiplication by the variable, and
 the one of ``p" (obtained by Fourier transform) is ``$-i\hbar\frac d{dx}$", the quantization of ``$qp$" is the quantum Hamiltonian:
\be\label{qp1}
H:=-\frac{i\hbar}2(x\frac d {dx}+\frac d {dx}x)
\ee
obtained form ``q" and ``p" by symmetrization.

An easy computation shows that the quantum flow is obtained by simple dilations:
\be
i\hbar\partial_t\psi=H\psi \Longleftrightarrow \psi^t(x)=e^{-t/2}\psi^0(e^{-t}x).
\ee
Therefore, if the initial condition is a coherent state ``pined up" at the origin, that is, $\psi^0(x)=(\pi\hbar)^{-1/4}e^{-\frac{x^2}{2\hbar}}$, we will have:
\be\label{qp2}
\psi^t(x)=(\pi\hbar)^{-1/4}e^{-t/2}e^{-\frac{e^{-2t}x^2}{2\hbar}}.
\ee
The state will start spreading as $t$ increases, and when $t\sim\frac 1 2 log(\frac 1 \hbar)$ the state is fully delocalized:
\be\label{qp3}
\psi^{\frac 1 2 log(\frac 1 \hbar)}(x)=\pi^{-1/4}e^{-\frac{x^2}2}.
\ee
This spreading phenomenon will give rise to a vanishing event as $t\sim log(\frac 1 \hbar)$ as:
\be\label{qp4}
\psi^{log(\frac 1 \hbar)}(x)=(\frac\hbar\pi)^{1/4}e^{-\hbar\frac{x^2}2}.
\ee
As $\hbar\to 0$ the state now is uniformly delocalized on the real line, that is on the unstable manifold of the classical flow issued from the origin.
\section{The case of the ``$8$"}
An emblematic example of hyperbolicity in Hamiltonian flows, although not stricto sensu ``chaotic', as having only one degree of freedom, is the case of an unstable fixed point. For example
the highest point of a double well potential. The ``physical"experience we have from instability is perfectly contained in this example. Suppose we have a little ball at the top
of a mountain: we know that, by getting a little kick it is going to fall down in the valley. We also know that, whatever  this kick is, as small as we want, it will fall down. But we know
also that if the kick is very small, the time the ball will take to start moving is going to be very long, a fact that is explained by reversing the time: if the ball comes from
the valley, with exactly the energy it takes to reach the top, it will take  an infinite time to arrive at the top. This is because near an hyperbolic fixed point the velocity
tends to $0$ as the particle tends to the fixed point.

Rephrased in more mathematical terms, the fixed point has a stable and an unstable manifold, and all the trajectories on them either end up at the fixed point, or ``come"
from it.

We will  consider \cite{TP1, TP3} the case of a quantum particle moving in a potential of the  form given in Fig.2:

\vskip 0.3cm

\centerline{\includegraphics{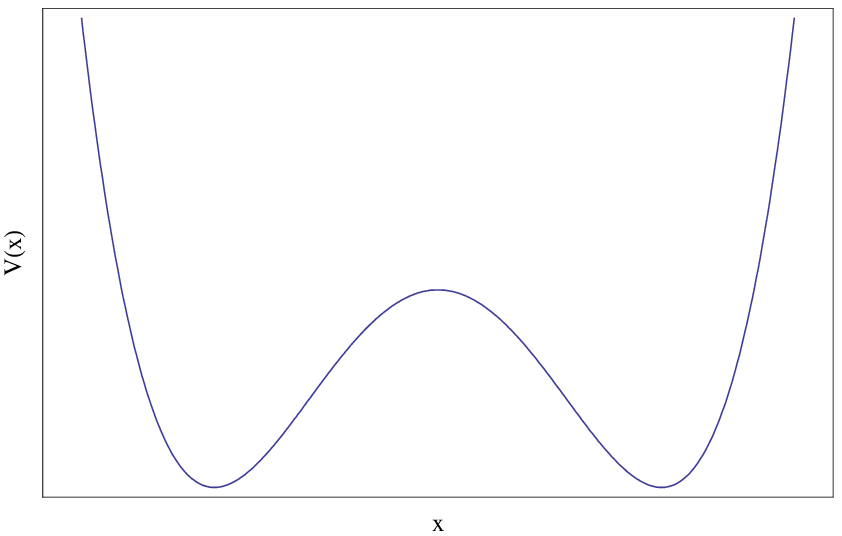}}

\centerline{Fig 2}

\vskip 0.3cm
for example
\be\label{81}
V(x)=x^2(x^2-1)
\ee
The trajectory, for the $0$ energy of the Hamiltonian $h(q,p)=\frac{p^2}2+V(q)$ is plotted in Fig. 3:

\vskip 0.3cm

\centerline{\includegraphics{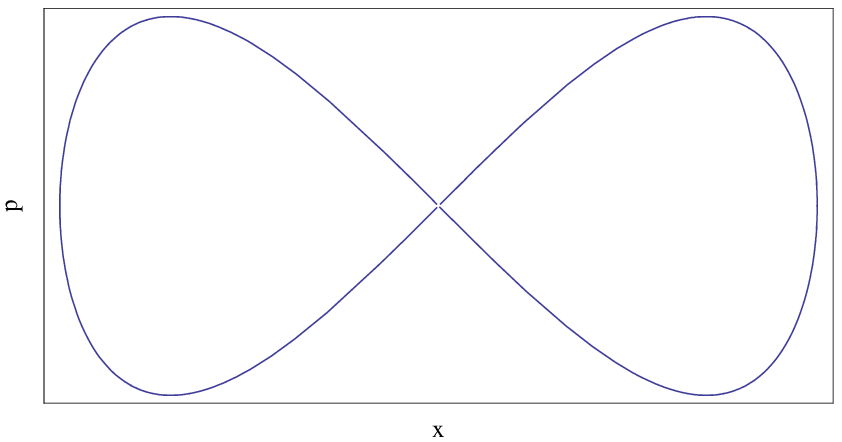}}

\centerline{Fig 3}

\vskip 0.3cm
It contains a fixed point at the origin, together with four branches corresponding to stable and unstable manifolds. The ``$8$" can be viewed as  a system similar to the
``dilation" case
in the preceding section, where the real lines corresponding to the stable and unstable manifolds have been glued at infinity. The quantum dynamics, in semiclassical
approximation,  can be described as follows: let us start, as in the preceding section, by a coherent state pinned up at the origin. Then the (semiclassical) evolution will behave
as follows:
\begin{itemize}
\item for $t$ small compared to $log(\frac 1 \hbar)$,  $\psi^t$ will remained localized at the origin
\item for $t=\frac 1 2 log(\frac 1 \hbar)$,  $\psi^t$ will be delocalized on half of the curve shown in the preceding figure.
\item for $t=log(\frac 1 \hbar)$,  $\psi^t$ will be relocalized at the origin
\item this qualitative evolution will be periodic with period $log(\frac 1 \hbar)$
\end{itemize}

\section{The semiclassical limit...not to classical mechanics}
We have seen in section \ref{scl} that, when expressed in a suitable way, the limit of the solution of the Schr\"odinger equation defines the classical flow. We mentioned already
also that this limit was not uniform with respect to the time scale. When we consider time longer than  a certain scale, $\hbar$-dependent, non classical effects are going to remain
at the limit $\hbar\to 0$. This is what we have seen in the case of  the $8$ 's case: the $log(\frac 1 \hbar)$ periodicity has no classical counterpart.

Another example \cite{TP2, TP3} is the case of the quantum evolution corresponding to a coherent state pinned up on a periodic trajectory of the classical flow which is linearly stable,
which means that, at the linear approximation, the neighborhood trajectories remain close to each other. In this case one shows that, for times of the order $\frac 1 \hbar$, the
coherent states, after a regime where it got delocalized like in the ``$8$" case, relocalizes.....at different points.

All these phenomena are traces of persistence of quantum effects at the classical limit.

\section{Discussion and general conclusion}
\subsection{Different scales of space and ``micro- versus macroscopic"}
It is generally believed that Quantum Mechanics is valid at microscopic scales, and classical one at macroscopic scale. Modern physics has changed this vision. Experimental
situations (Rydberg atoms), where highly excited states are created, show  single atoms of ``size" of the order of small bacterias ($\sim 1 \mu m$). Simple quantum systems at semiclassical regimes
are nowadays frequently prepared, and remain excited for time of the order of several seconds. One even thinks that atoms of macroscopic size could be present in the Universe.
Therefore the dichotomy micro/macro is not anymore in correspondence with the quantum/classical classification.

\subsection{Different scales of time and the deconstruction of the infinite}\label{sca}
The discussion before shows clearly that different scales of time exhibit different behaviour at the semiclassical limit. In the example of the ``$8$" we get:

\begin{itemize}
\item $0\leq t<<\frac 1 2 \log(\frac 1 \hbar)$,  (semi)classical regime, the classical flow is recovered
\item $t\sim  \frac 1 2 \log(\frac 1 \hbar)$,  the ``point" has disappeared, delocalization
\item $t\sim   \log(\frac 1 \hbar)$,  the ``point"  reappears, due to quantum effects.
\end{itemize}


All these scales of time disappear at the classical limit $\hbar \to 0$ and ``pass" in the infinite time of Classical Mechanics.

\subsection{Invariants, ontology and what we see}

It is a general fact that Quantum Mechanics destroys our classical way of considering space. In particular it is not anymore possible to assign to an electron a trajectory, made by
a succession of points in space. Coherent states provide the best possible localization (of size $\hbar$): one has to imagine coherent states as associated to a little ball of
radius $\sqrt\hbar$. But as we have seen in the previous sections, when the state evolves for a time of order $log(\frac 1 \hbar)$ this ball is dilated and delocalizes on a
manifold. Not any manifold but some invariant manifold of the classical flow. Those manifolds are crucial tools of Classical Mechanics, but in the classical paradigm they are only
{\it tools}\rm. Quantum Mechanics gives them another status, actually the same as the point: Quantum Mechanics transforms points into invariant manifolds. 

\subsection{Conclusion: undeterminism versus unpredictability}

Non-deter-\\minism has often been  considered as  a problem of Quantum Mechanics. The result of a quantum measurement process is fully, intrinsically, random, and it follows a
probability law, as we explained earlier. In the case of a measurement of the position  done at time $t$, the probability of finding  the result ``$x$" is given precisely by
$|\psi^t(x)|^2$.

At the contrary Classical Mechanics is  a fully deterministic theory in the sense that the equations
of motion have, for any time, a unique solution. The theory of dynamical systems, especially in the chaotic situation, has brought, with the
phenomenon of sensitivity to initial conditions, another element of consideration. Although fully deterministic a chaotic dynamical system has a peculiar behaviour: it is hardly predictable. More precisely the
knowledge of the initial condition necessary for a good knowledge of the solution for long time has to be extremely precise, and therefore difficult to handle in realistic
situations. 

At the mathematical level we have seen that the unstable manifold $\Lambda$ of a fixed point $x_0$ is the set of points whose trajectories end up at $x_0$ at $t=-\infty$.
\be\label{conc1}
\Phi^{-\infty}(\Lambda)=\{x_0\}
\ee
Therefore this limiting procedure $t=-\infty$ suggests that it should exist a ``flow" $\Phi^{+\infty}$ obtained by taking the inverse of (\ref{conc1}):
\be\label{conc2}
\Phi^{+\infty}(\{x_0\})=\Lambda.
\ee
The meaning of this should be that $\Phi^{+\infty}x_0$ can be any point $y\in\Lambda$. 

In other words the infinite time limit of the flow, $\Phi^{+\infty}$, should be non
deterministic. This is in a certain way the morality of the sensitivity to initial condition: {\it for any perturbation of the initial condition there is a time for which the
evolution reach any point of the unstable manifold}\rm. It is enough to consider the dilation case to be convinced:
\be
\Phi^t (x)=e^t x.
\ee
Take any point on the unstable manifold $\Lambda=\{(y,0)\}$: the point $(e^{-t}y,0)$ end up at time $t$ at $(y,0)$.

Therefore we see that a formal limit $t=\infty$ in classical mechanics replaces  unpredictability by undeterminism, but only ``formally", without true mathematical structure,
since the equations (\ref{conc1}, \ref{conc2}) are only ``images": in Classical Mechanics the flow is defined at any time $t$, but the time $t=\infty$ can be only approached
asymptotically, and this is  the subject of the theory of dynamical systems.

We have seen that in Quantum Mechanics the situation is, in a certain sense, opposite: the flow on space doesn't exist, but the space can be recovered by a measure 
of the position
quantity $x$. And we have seen that, in the semiclassical setting, the long time dispersion of the wave function exhibits exactly the unstable manifold as possible values for the
result of the measurement: the semiclassical limit of long time evolution gives a precise mathematical sense to the formal classical construction we just described and,
moreover, gives a deconstruction of this  temporal point at infinity, by showing off different scales as expressed in section (\ref{sca}).

\vskip 0.5cm

Sensitivity to initial conditions gives rise, in Classical Mechanics when
considering infinite time evolution, to a certain form of undeterminism, and  this undeterminism overlaps exactly with the quantum one in the semiclassical limit. Maybe is it
time to start to think that, instead of complaining about randomness in Quantum Mechanics, one should reconsider, and abandon as infinite time evolutions are concerned, the strict classical determinism.


\end{document}